\documentclass[preprint,12pt]{elsarticle}
\biboptions{sort&compress}

\usepackage[utf8]{inputenc}
\usepackage{multirow}

\usepackage{graphicx}
\usepackage{amssymb}

\journal{Parallel Computing}

\begin{document}

\begin{frontmatter}

\title{On the Impact of the Migration Topology\\on the Island Model}

\author[act]{M. Ruci\'{n}ski\fnref{fn-marek}}
\ead{marek.rucinski@plymouth.ac.uk}
\author[act]{D. Izzo\corref{cor-dario}}
\ead{dario.izzo@esa.int}
\author[act]{F. Biscani}
\ead{francesco.biscani@esa.int}

\address[act]{Advanced Concepts Team, European Space Agency (DG-PI), Keplerlaan 1, 2201 AZ Noordwijk, The Netherlands}
\fntext[fn-marek]{Present address: Centre for Robotics and Neural Systems, University of Plymouth, Drake Circus, Plymouth, PL4 8AA, United Kingdom}
\cortext[cor-dario]{Principal corresponding author. Phone: +31(0)71 565 3511, Fax: +31(0)71 565 8018}

\begin{abstract}
Parallel Global Optimization Algorithms (PGOA) provide an efficient way of~dealing with hard optimization problems.
One method of~parallelization of~GOAs that is frequently applied and commonly found in the contemporary literature is
the so-called Island Model (IM). In~this paper we analyze the impact of the migration topology on the performance of~a~PGOA
which uses the Island Model.
In particular we consider parallel Differential Evolution and Simulated Annealing with Adaptive Neighborhood and~draw
first conclusions that emerge from the conducted experiments.
\end{abstract}


\end{frontmatter}

\section{Introduction}
\label{sec:intro}

The Island Model \cite{Belding95} is a coarse-grained approach to the parallelization of Global Optimization heuristics. 
Originally developed for Genetic Algorithms and inspired by the theory of punctuated equilibria \cite{Cohoonetal87}, it can 
actually be implemented also for algorithms based on drastically different paradigms such as Particle Swarm 
Optimization or Simulated Annealing. For Parallel Genetic Algorithms it has been shown, that employing the Island Model
leads to increased algorithm performance \cite{CantuPaz00} what can be explained in terms 
of improved balance between exploitation and exploration of the solution space. In the Island Model each island can exchange information with its 
neighbor island as defined in the graph of possible inter-island links commonly referred to as migration topology. 
With the increase in CPU number and power, the number of islands contributing to one optimization can grow significantly 
and the resulting optimization is thus affected more clearly by the way the information is exchanged between the islands 
and in particular by the migration topology. Broadly speaking, the migration topology has two effects on the underlying 
optimization process. The first one, beneficial, is the super-linear speed-up caused by the information exchange, and the 
second one, being sometimes an issue, is the BUS and CPU overhead caused by the required information flow. In this paper we 
consider fourteen different migration topologies, some studied here in this context for the first time, and we present an 
extensive study on their effect on the performances of two quite popular Global Optimization Algorithms, Differential 
Evolution \cite{StornPrice97} and Simulated Annealing in its variant with adaptive neighborhood proposed by Corana et al. \cite{Coranaetal87}. We 
consider topologies with a large number of nodes (namely up to 1024) filling what we felt an existing gap in the 
literature on the Island Model for Global Optimization Algorithms. In this work we do not consider the cost of the 
CPU/BUS overhead introduced by the adopted migration scheme as we take the standpoint of studying the overall algorithm 
performance in an ideal case where sufficient amount of information (i.e. migrating solutions) can flow within the
network without significant costs.

The paper is organized as follows. We start with a brief overview of up-to-date research on the migration topology in the
Island Model. Then our experimental methodology is presented, including descriptions of the used benchmark problems,
the tested algorithms, and the procedures of data analysis. We continue with a detailed investigation of obtained results
and their interpretation, finishing with a brief summary and description of future research.

\section{Related Research}
\label{sec:relatedresearch}

In the light of the fact that the Island Model was first introduced in the connection with Parallel Genetic Algorithms, 
it is not surprising that it is in this context that all fundamental issues involved, including the migration topology, received 
the most attention. Numerous reviews can be found, published regularly since mid-90's, where the progress in this field is 
well documented (see for instance \cite{CantuPaz95, Schwehm96, CantuPaz98, AlbaTroya99, Konfrst04}). Initially, the 
impact of the migration topology on the overall performance of the algorithm has been ``traditionally neglected'' 
\cite{CantuPaz95} and its choice was dictated mainly by the native topology of the used computing machine. As the research in the field
progressed, a number of experimental studies aiming to decide on the best topology for a coarse-grained Parallel Genetic
Algorithm appeared (like for example \cite{GordonWhitley93, CantuPazMejiaOlvera94}), but also some theoretical attempts
to tackle the problem have been performed. Among these one of the most fruitful was the dissertation of Cant\'{u}-Paz 
\cite{CantuPaz00}. While it provided a lot of theoretical insight into PGAs in general, especially on the importance of
the population and sub-population sizing, it contained also significant results in the context of the analysis of the
migration topology. Namely, in this work it has been proved that, under certain assumptions (including neglecting the
communication costs), the fully-connected topology is the optimal choice for a PGA. For situations where the
communication costs are significant, the author provided formulas to calculate the optimal network connectivity. Apart
from the theoretical analysis, the study also included their experimental verification using synthetic objective functions.

The idea of a coarse-grained parallelization approach was of course not studied exclusively in connection to Genetic Algorithms. Very
similar concepts have either appeared simultaneously or have been adopted from other fields resulting in a multitude of
parallel versions of different metaheuristics like Simulated Annealing, Tabu Search, GRASP, Ant Colony algorithms and
Genetic Programming to name at least few (see for instance \cite{RibeiroHansen02, Alba05} for excellent reviews,
see \cite{Fernandezetal03} for an empirical study of multipopulation Genetic Programming, which also looked at the
migration topology issue). Because all those implementations shared many concepts, soon a general term of Parallel
Cooperative Metaheuristics \cite{RibeiroHansen02} has been forged.

The Island Model as used in the context of this paper certainly belongs to the broader family of multiple-walk 
algorithms with cooperative threads \cite{RibeiroHansen02}. Our consistent use of the narrower term is motivated by the 
fact that we consider exclusively algorithms in which the information exchanged between concurrent optimization 
processes consists of the problem solutions (chromosomes, decision vectors) only. This makes the idea directly 
applicable to a great number of algorithms (population-based as well as Local Search-based), which is not the case for 
the dedicated approaches (like tabu lists sharing in Parallel Tabu Search). Moreover, such specification of the 
parallelization paradigm allows for easy introduction of cooperative heterogeneous approaches, where different islands use different 
optimization algorithms. 

Finally it may be worth noting that our study is certainly not the first one in which the Island Model-like
parallelization is applied to the Differential Evolution and Simulated Annealing algorithms. Both these ideas have been
successfully implemented to date (see for instance \cite{Tasoulisetal04, KwedloBandurski06, Laursen94, Izzoetal09}).
The unique contribution of this study is the first systematic investigation of the influence of the migration topology,
being one of the Island Model parameters, on the global performance of the parallelized algorithm on a limited set of
representative problems, in the context of applications in Global Optimization.

\section{Experimental Methodology}
\label{sec:methodology}

In this section we provide a detailed description of the experimental set-up which has been used in our research. We
start with describing the Global Optimization problems which we used as a test-bed to compare the performance of tested
algorithms. The significant computational cost of the performed numerical experiments prevented us from using standard
test sets for Global Optimization Algorithms and forced us to select a limited number of widely used problems.
We then present the analyzed Global Optimization Algorithms and their parameters -- both algorithm-specific and related to
the Island Model. Large part of the section is devoted to describing the tested migration topologies. We continue with
giving the details about what data have been collected from the experiments and how they were analyzed. Finally, we
provide technical information about software and hardware which have been used to conduct the experiments.

\subsection{Global Optimization Problems}

The following Global Optimization problems have been used in our research as a small benchmark set for tested algorithms.

\subsubsection{Rastrigin Function Minimization}

The Rastrigin function is given by the following equation:
\begin{equation}
	f_R(\textbf{x}) = 10 n + \sum_{i=1}^n \Big(x_i^2 - 10 \mathrm{cos}(2 \pi x_i)\Big)
\end{equation}
where $n$ denotes the number of dimensions. Rastrigin function is commonly used in benchmark sets for comparison
of performances of optimization algorithms. The function is multimodal and separable. It has the global minimum
at 
\begin{equation}
	f_R(\textbf{x}^\star) = 0
\end{equation}
in
\begin{equation}
	\textbf{x}^\star = [ 0, 0, \ldots, 0 ]
\end{equation}
For the purposes of this research, we used a function with $n = 250$ dimensions and a solution space
bounded by  $-5.12 \leq x_i \leq 5.12$.

\subsubsection{Schwefel Function Minimization}

The Schwefel function is given by the following equation:
\begin{equation}
	f_S(\textbf{x}) = 418.9829 n + \sum_{i=1}^n x_i \mathrm{sin}\Big(\sqrt{|x_i|}\Big)
\end{equation}
where $n$ is the number of dimensions. Schwefel's function is multimodal and separable. The function has the
global minimum at
\begin{equation}
	f_S(\textbf{x}^\star) = 0
\end{equation}
in
\begin{equation}
	\textbf{x}^\star = [ -420.9687, -420.9687, \ldots, -420.9687 ]
\end{equation}
For the purposes of this research, we used a function with $n = 250$ dimensions and a solution space bounded by $-500 \leq x_i \leq 500$.

\subsubsection{Lennard-Jones Potential Minimization}

The Lennard-Jones problem has been introduced in \cite{Lennard-Jones31} and concerns finding positions of $N$ atoms in a
cluster which result in a minimal total Lennard-Jones potential, which is expressed as:
\begin{equation}
	V = 4 \sum_{i=1}^{N-1} \sum_{j=i+1}^{N} \Big( \frac{1}{r_{ij}^{12}} - \frac{1}{r_{ij}^6}\Big)
\end{equation}
where $r_{ij}$ is the Euclidean distance between $i$-th and $j$-th atom. Positions of atoms are expressed in Cartesian
coordinates and when one prunes out redundant solutions by using fundamental symmetries the number of problem
dimensions becomes $n=3N-6$. For our purposes we used the Lennard-Jones potential minimization problem in a cluster of
$31$ atoms, which resulted in a problem with $87$ dimensions. Optimal solution of the problem for an arbitrary number
of atoms is not known. The state-of-the-art results for clusters of different numbers of atoms, including 31, can be
found for instance in \cite{WalesDoye97} and the value
\begin{equation}
	V_{31}^\star = -133.586422,
\end{equation}
first reported in \cite{Northby87}, has been used for assessment of the quality of obtained solutions. The solution space
bounds considered are $-3 \leq x_i \leq 3$.

In order to increase the readability, in the next sections of this paper we use the following designations when
referring to the above optimization problems in tables and figures: Rastrigin function minimization with 250 dimensions
-- $f_R$, Schwefel function minimization with 250 dimensions -- $f_S$, Lennard-Jones potential minimization with 31
atoms -- $V_{31}$.

\subsection{Base Algorithms}

The description of the parallel Global Optimization Algorithms considered here can be
decomposed into two parts: the description of sequential algorithms being parallelized with their specific parameters
(we refer to them as the \emph{base algorithms}) and the description of the parallelization strategy.

\subsubsection{Base Algorithm 1 - Differential Evolution}

The Differential Evolution algorithm has been originally proposed by R.~Storn and~K.~Price in~\cite{StornPrice97}.
It is population- and generation-based, and defines mutation, crossover and selection operators (which should not be 
confused with identically called operators defined for Genetic Algorithms).
In the paper authors propose various alternatives for mutation and crossover, and by picking different options for each
of them one can construct different variants of the algorithm. Using the notation introduced by the authors in the same
publication, the algorithm variant used during our study is denoted as $\mathrm{DE}/best/2/exp$, which means using
the best solution in the population as the mutated vector, two difference vectors and the exponential crossover scheme.
Apart from these three components, the DE algorithm has the following parameters: population size $NP$, amplification
factor $F$ and crossover constant $CR$. For our study we fixed values of these parameters to be $NP = 20$, $F = 0.8$ and
$CR = 0.8$. 

\subsubsection{Base Algorithm 2 - Simulated Annealing with Adaptive Neighborhood}
\label{sec:methodology-algorithms-asa}

The second parallelized algorithm is a variant of Simulated Annealing proposed by Corana et al. in \cite{Coranaetal87}
which is an adaptation of the original combinatorial Simulated Annealing algorithm formulated in \cite{Kirkpatricketal83}
for optimization of functions defined in the continuous domain. A characteristic component of Corana's version is the
adaptive adjustment of the solution neighborhood based on the number of solutions accepted by the algorithm in each
search direction during the sampling period. This neighborhood adjustment is a way to ensure that the algorithm
adapts to the properties of the optimized objective function, preserving the balance between global and local search.

Our implementation of this variant of Simulated Annealing uses re-an\-neal\-ing and has the following parameters: starting
and final temperatures $T_s$ and $T_f$, the sample size for the neighborhood adjustment $N_s$ (in the original paper
called the ``number of cycles''), the number of neighborhood adjustments on one temperature level $N_T$ (``number of 
step adjustments'' in the original paper) and the initial step size $v_0$. The number of temperature adjustments 
performed throughout the algorithm is determined automatically based on the number of objective function evaluations 
available before starting a new re-an\-neal\-ing cycle and the number of dimensions of the objective function. Intermediate 
temperatures are calculated so that they follow a geometric series falling from $T_s$ to $T_f$ in the given number of 
steps. 

On the set of considered test problems the Simulated Annealing algorithm performance is more sensitive to the values of 
its parameters than Differential Evolution. Thus it has been decided to check for this algorithm if fine-tuning the 
parameters affects significantly the performance of compared parallelization strategies and, in consequence, obtained 
results. First, computations for all test problems were performed using the rule-of-thumb values, which are the 
following: $T_s = 1.0$, $T_f = 0.001$, $N_s = 20$, $N_T = 1$ and $v_0 = 1.0$. The parameters of Simulated Annealing were 
then tuned manually for each test problem separately, based on the progress of optimization of a single instance of the 
algorithm, until smooth convergence toward global optimum was observed, without having the algorithm stuck in one of 
the first encountered local minima. Chosen parameters may not be optimal, but for some of the test problems 
(particularly for the Schwefel function) lead to significant improvements in the quality of the final solution when 
compared to the instance of the algorithm with untuned parameters. Tuned values of the Simulated Annealing algorithm
parameters for considered test problems are reported in table \ref{tbl:asaparams}. 

\begin{table}
\centering
\begin{tabular}{cccccc}
Problem & $T_s$ & $T_f$ & $N_s$ & $N_T$ & $v_0$ \\
\hline
$f_R$ & $0.1$ & $0.0001$ & $5$ & $2$ & $1.0$\\
$f_S$ & $1.0$ & $0.001$ & $10$ & $1$ & $1.0$\\
$V_{31}$ & $1.0$ & $0.0001$ & $10$ & $2$ & $1.0$\\
\end{tabular}
\caption{Manually tuned parameters of the Simulated Annealing algorithm for the test problems.}
\label{tbl:asaparams}
\end{table}

In order to increase the readability, in the next sections of this paper we use the following designations when 
referring to the above optimization algorithms in tables and figures: described variant of Differential Evolution -- 
$DE$, Simulated Annealing with Adaptive Neighborhood with untuned parameters -- $SA_u$, Simulated Annealing with 
Adaptive Neighborhood with tuned parameters -- $SA_t$. 
 
\subsection{Island Model Parameters}
\label{sec:methodology-island-model-params}

Parallelizing a Global Optimization Algorithm using the Island Model involves taking decisions about quite a few
model's parameters. As summarized in \cite{CantuPaz00}, these parameters are the following:
\begin{itemize}
	\item The \emph{number of islands} representing the number of instances of the base algorithm (separate populations) existing in parallel;
	\item The \emph{migration topology} determining feasible migration paths;
	\item The \emph{migration rate} telling how many individuals migrate from the source population at a time;
	\item The \emph{migration frequency} telling how often the migration occurs;
	\item The \emph{migration algorithm} specifying all remaining details.
\end{itemize}
The last parameter needs a more extensive description.

First of all, migration can be \emph{synchronous} or \emph{asynchronous}. In the case of synchronous migration the 
exchange of individuals occurs at the same time for all islands, and during that time no computation is performed. In 
the asynchronous case each island ``delegates'' migrating individuals as soon as it is ready to do so, without taking 
into account the state of other islands. There are two main factors that influence the choice of one variant over the 
other. First, in a distributed computing environment synchronizing the execution of all islands may be expensive, and
may require a centralized approach to control computations which in turn affects scalability.
Moreover, global synchronization has a negative impact on the parallel speed-up when islands 
are run on heterogeneous computing machines (faster processors will be idle while waiting for the slowest one to complete). Asynchronous model does not suffer from these conditions, especially if 
communication between two islands is also implemented in an asynchronous way (i.e. the computation on the target island 
is not suspended by the communication). On the other hand, presence of global synchronization mechanism makes such 
parallel algorithm much easier to monitor and control than its asynchronous counterpart. The (realistic) assumption of 
stochastic computation time does not affect the flow of the synchronous parallel algorithm and a deterministic 
algorithm is possible to obtain. For the asynchronous parallelization in turn, non-determinism of the execution times implies
non-determinism of communication leading to the non-determinism of the whole algorithm, even if stochastic properties
of the base algorithms (DE or SA in our case) are controlled using a seeded pseudorandom number generator. In addition, the lack of global
synchronization between the parallel islands makes it much harder to define the global state of the algorithm at an
arbitrary point of time, and thus creates problems with monitoring the algorithm for instance for the purpose of
implementing certain stopping conditions (like reaching a particular value of the objective function by any island).

The next detail that requires specification when formulating the migration algorithm is the way of selection of
individuals that are supposed to migrate from the source population and how incoming individuals are inserted to the 
destination population. The spectrum of possible options is very well described in \cite{Schwehm96} and in the most 
cases can be reduced to a choice between randomness and elitism.

The last choice that can be ascribed to the migration algorithm details is the method of distribution of the migrating
individuals. Two alternative choices can be ``point-to-point'' migration (where individuals from a source island end up in
exactly one destination island, being the neighbor of the source island as defined by the migration topology) versus
``broadcasting'' migration (where individuals from a source island are sent to all neighbor islands). The communication
method may be linked with the algorithm of selection for migration mentioned above -- for example a possible choice is,
instead of choosing a limited subset of individuals to migrate, to split the whole population in such a way that each
neighbor island receives a distinct piece of it (this way of communication is one of the core assumptions under which
theoretical calculations for Parallel Genetic Algorithms are performed in \cite{CantuPaz00}).

As for the other parameters of the Island Model, migration rate and migration frequency are unavoidably linked to the 
properties of the algorithm being parallelized. Migration rate is obviously limited by the population size used by the 
algorithm (which can be 1 in the limiting case). Migration frequency may be limited by the time (possibly measured as 
the number of performed objective function evaluations) the algorithm needs to make a sensible improvement to the 
population. ``Flooding'' a population with foreign individuals with too high frequency may make the algorithm lose its 
properties and simply stop working as expected \cite{Tanese89}. Usually it is desirable to allow an island to 
perform a certain amount of work before the migration occurs -- one can even allow migration only after an island 
gets stuck in a local minimum \cite{Braun91}. 

In the following paragraphs we present choices made for the Island Model parameters presented above and explain
the rationale behind taken decisions.

\subsubsection{Number of Islands}
\label{sec:methodology-no-of-islands}

In the experiments, four distinct numbers of islands have been considered: $128$, $256$, $512$ and $1024$. These figures 
are larger than values commonly found in the up-to-date literature treating about the Island Model and its 
implementations (in which the number of 64 nodes is not exceeded) and there are several reasons for this. The first is 
the increasing availability of distributed computing environments which do not suffer from strict limits on the number 
of processors involved in the computation as it was the case with parallel systems which were in use when first major 
works on parallel Global Optimization Algorithms (mainly Parallel Genetic Algorithms) have been done. This increase in 
the number of CPUs available to a typical researcher is coupled with the increase in communication speed between 
computers which makes it feasible to implement Island Model-based computations with larger and larger numbers of 
islands. The second reason why such large numbers of islands are considered in this paper is the impact of this 
parameter on the main object of this study -- that is on the migration topology. When the number of islands in the model 
grows, both introduction of new topologies becomes possible (namely topologies which do not manifest their intrinsic 
properties if the number of nodes is not sufficiently high) and distinct properties of the 
well-established ones become more emphasized. An example of such an emphasis is given in table \ref{tbl:fc-hcratio} 
where the ratio of the number of edges in the Fully Connected topology to the number of edges in the Hypercube topology 
with the same number of nodes is compared for two different numbers of nodes: 8 (values of this order of magnitude are 
most commonly found in contemporary literature) and 1024. It is obvious that if that parameter has any impact on the 
global performance of the algorithm, this fact will be much more evident in the latter case, as the ratio is two orders 
of magnitude higher than in the former case. 

The choice of the numbers of islands being powers of 2 has been driven by the fact that for such numbers of nodes it
is possible to construct a full hypercube of an appropriate dimension.
\begin{table}
\centering
\begin{tabular}{cccc}
Number & \multicolumn{2}{c}{Number of edges} & \\
of nodes & Fully Connected & Hypercube & Ratio\\
\hline
$8$ & $56$ & $24$ & $\mathbf{2.33}$\\
$1024$ & $1047552$ & $10240$ & $\mathbf{102.3}$\\
\end{tabular}
\caption{Example of a topology property being emphasized with the growing number of islands.}
\label{tbl:fc-hcratio}
\end{table}

\subsubsection{Migration Topology}

We have considered the total of 14 different migration topologies in our study. Most of these have been
taken from the up-to-date literature on the Island Model and parallel optimization in general.
In this section we briefly describe each tested topology starting from the most simple ones, mentioning some of their
major properties and, where possible, providing references to related papers. Some of the most important parameters of
considered migration topologies can be found in table \ref{tbl:topology-params} at the end of this section.

\begin{figure}
\centering
	\begin{tabular}{cc}
		\includegraphics[keepaspectratio=true,width=0.3\linewidth]{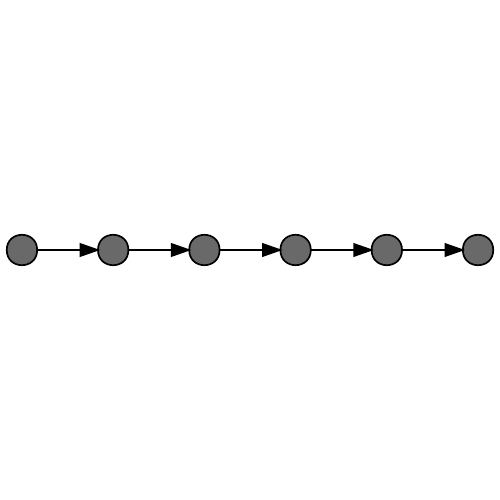}
		&
		\includegraphics[keepaspectratio=true,width=0.3\linewidth]{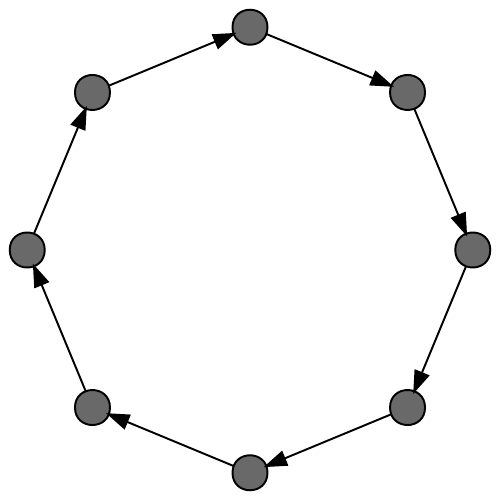}\\
		(a) & (b)\\
	\end{tabular}
   \caption{Chain (a) and Uni-directional Ring (b) topologies.}
\label{fig:topo-chain-one_way_ring}
\end{figure}

One of the most simple topologies one can imagine is a \emph{Chain topology} (figure \ref{fig:topo-chain-one_way_ring}a, 
also known as \emph{line topology} or \emph{line network}) where islands are organized in a sequence and the communication 
is allowed only between neighbor islands. While not very popular in the context of the Island Model, this kind of 
topology can be often found in studies on distributed algorithms, for example on the complexity of distributed sorting 
algorithms \cite{Sasaki02}. In the Chain topology communication may be either uni- or bi-directional. We decided to 
consider the former case, as it possesses a property which no other considered topology has: exactly one island is not 
affected by the communication -- the first island in the chain does not receive any foreign individuals. 

\begin{figure}
\centering
	\begin{tabular}{ccc}
		\includegraphics[keepaspectratio=true,width=0.3\linewidth]{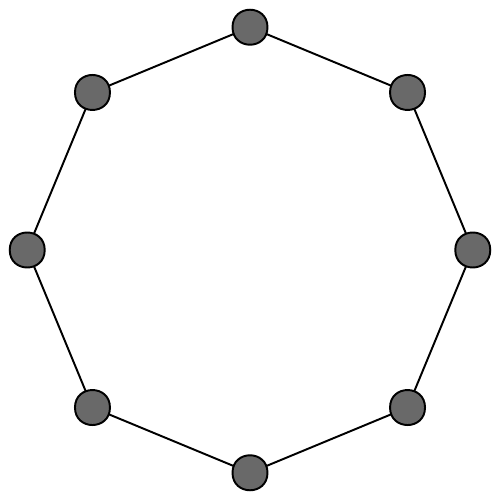}
		&
		\includegraphics[keepaspectratio=true,width=0.3\linewidth]{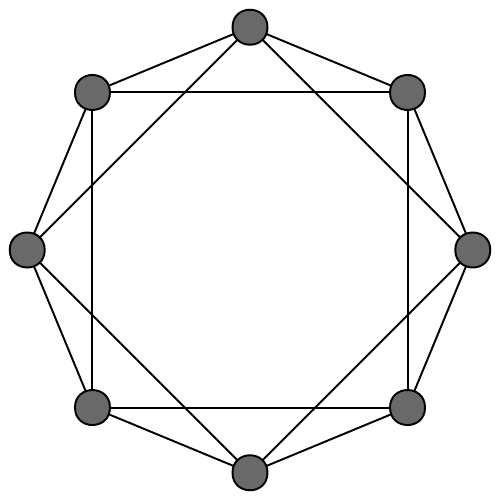}
		&
		\includegraphics[keepaspectratio=true,width=0.3\linewidth]{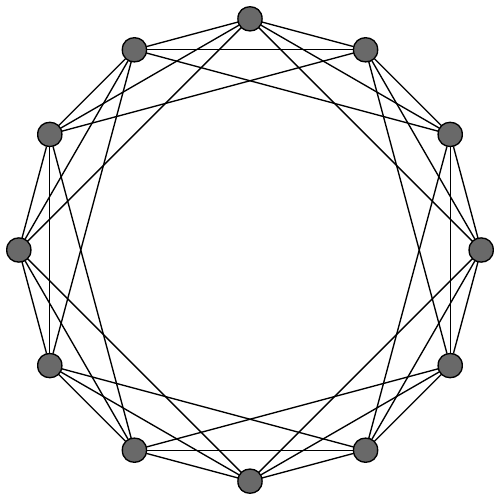}\\
		(a) & (b) & (c)\\
	\end{tabular}
   \caption{Ring (a), Ring+1+2 (b) and Ring+1+2+3 (c) topologies.}
\label{fig:topo-ring-ring12-ring123}
\end{figure}

When one connects the last and first element of the Chain topology, one obtains \emph{Uni-directional Ring topology}. 
(figure \ref{fig:topo-chain-one_way_ring}b). Of course, in this case the communication may be also uni- or 
bi-directional, and we include both variants in our experiments. We refer to the bi-directional version simply as the 
\emph{Ring topology} (figure \ref{fig:topo-ring-ring12-ring123}a). Both variants of this topology, are very commonly 
used both in the context of traditional GA Island Model, and in parallel optimization in general. Studies which include 
some variants of the topology in different contexts can be found for example in \cite{Homayounfaretal03, 
Starkweatheretal91, GordonWhitley93, CantuPaz00, CantuPazMejiaOlvera94, Izzoetal09, Fernandezetal03}. Sometimes, the Ring topology is 
referred to as \emph{Stepping Stone Model}, we prefer however to refrain from using this term because of its unclear 
character. 

A simple extension to the Ring topology can be proposed, using which a series of topologies with gradually decreasing
diameter and increasing degree of connectivity can be obtained. Namely, one can add edges that connect every second island
in the basic ring, every third island and so on. First two topologies from this series have been included in our experiments.
We called them \emph{Ring+1+2} and \emph{Ring+1+2+3 topology} (figure \ref{fig:topo-ring-ring12-ring123}b and c).
Similar concept can be found in \cite{CantuPaz00}, where different topologies having the same degree of connectivity
have been analyzed in the context of their performance in island-model parallel GA.

\begin{figure}
\centering
	\begin{tabular}{ccc}
		\includegraphics[keepaspectratio=true,width=0.3\linewidth]{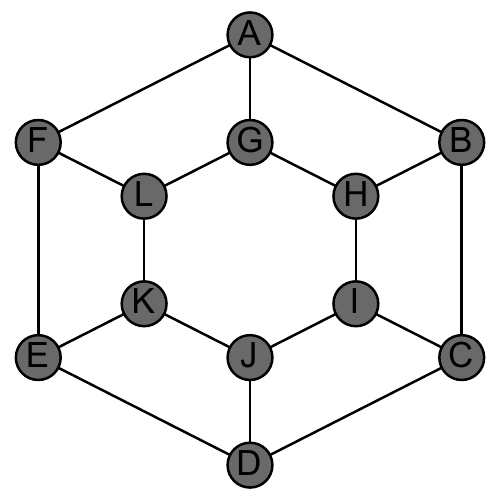}
		&
		\includegraphics[keepaspectratio=true,width=0.3\linewidth]{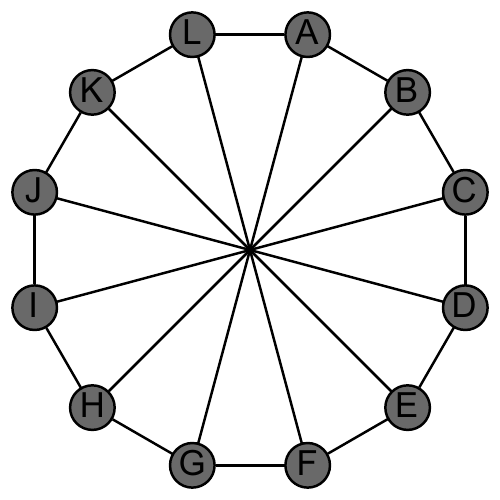}
		&
		\includegraphics[keepaspectratio=true,width=0.3\linewidth]{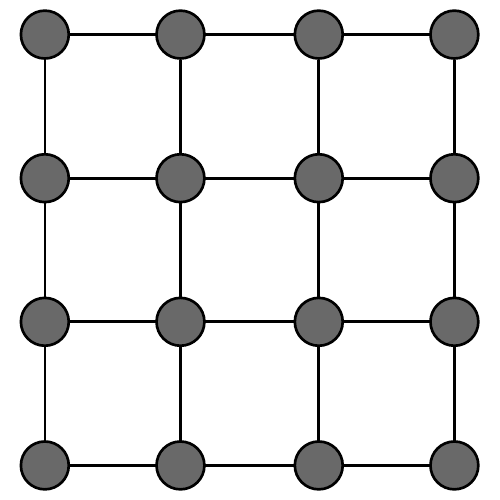}\\
		(a) & (b) & (c)\\
	\end{tabular}
   \caption{Torus (a), Cartwheel (b) and Lattice (c) topologies.}
\label{fig:topo-torus-cartwheel-lattice}
\end{figure}

Probably the second most popular migration topology in the literature about the Island Model is the \emph{Torus 
topology} (figure \ref{fig:topo-torus-cartwheel-lattice}a). This popularity is at least partially due to the fact that 
in early parallel computing systems with distributed memory (like transputers) this was often the default layout of 
hardware connections between processing units (and thus the most robust one in terms of communication delays). The Torus 
topology appears however in many variants, differing with allowed paths of communication (certain edges may be 
uni-directional while others bi-directional) and the number and arrangement of ``rings'' from which the topology is 
built. We decided to implement the most generic variant of the topology, i.e. the one in which islands are organized in 
two parallel rings with corresponding islands connected, and with all communication links bi-directional. Similar 
structure, but with uni-directional links along the rings is sometimes called \emph{Ladder topology}. Example papers 
mentioning the Torus (or Ladder) topology are \cite{Muehlenbeinetal91, BianchiniBrown93, CantuPazMejiaOlvera94, 
CantuPaz00}. 

The \emph{Cartwheel topology} (figure \ref{fig:topo-torus-cartwheel-lattice}b), mentioned for instance in \cite{CantuPaz00}, is a 
ring with additional edges connecting all pairs of islands laying on opposite ``ends'' of the basic structure. Although 
not evident at first glance, this topology is nearly identical to the Torus topology variant used in this paper. 
Comparing connections between corresponding nodes in both topologies (see figure \ref{fig:topo-torus-cartwheel-lattice}a 
and \ref{fig:topo-torus-cartwheel-lattice}b) it is easy to notice that when links F-A and L-G in the Torus topology 
graph are replaced with L-A and F-G, one obtains a graph that is homomorphic to the one of the Carthweel topology. For 
any number of islands, these two topologies always differ only with two edges. This small difference has however an 
impact on the diameter of the resulting network, which is slightly smaller for the Cartwheel topology for numbers of 
islands being multiplies of 4 (see table \ref{tbl:topology-params}), and that is why we decided to include the Cartwheel 
topology in our analysis. 

While not necessarily popular in the literature concerning the Island Model, \emph{Lattice topology} (figure 
\ref{fig:topo-torus-cartwheel-lattice}c) can be found in a number of studies on parallel GA in the context of 
fine-grained parallelization (see for instance \cite{Cohoonetal87, GordonWhitley93, Fernandezetal03}). It can also be found in older 
studies on parallel algorithms as this kind of topology was usually very easy to obtain (or the default available) on 
already mentioned transputer systems. When one is free from hardware constraints, it is of course possible to construct 
rectangular Lattices of arbitrary width and height. However, in order not to introduce additional parameters, we 
considered only square lattices. In two cases (for 128 and 512 islands) this led to obtaining topologies that were not 
purely rectangular (one row/column filled only partially), but there was no easy remedy for this issue because of the 
constraints introduced by the Hypercube topology, described in the next paragraph. 

\begin{figure}
\centering
	\begin{tabular}{ccc}
		\includegraphics[keepaspectratio=true,width=0.3\linewidth]{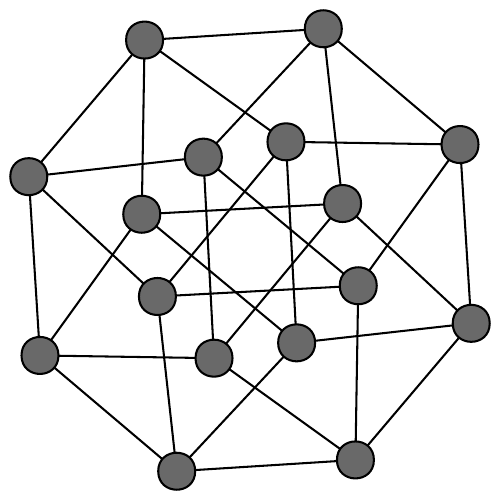}
		&
		\includegraphics[keepaspectratio=true,width=0.3\linewidth]{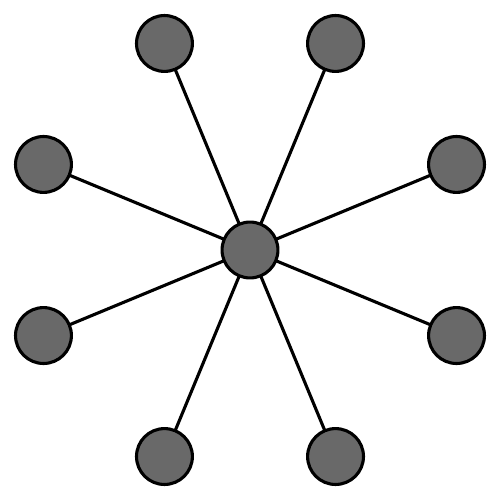}
		&
		\includegraphics[keepaspectratio=true,width=0.3\linewidth]{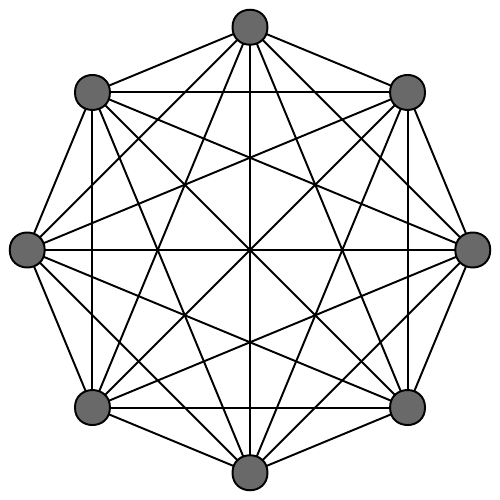}\\
		(a) & (b) & (c)\\
	\end{tabular}
   \caption{Hypercube (a), Broadcast (b) and Fully Connected (c) topologies.}
\label{fig:topo-hypercube-broadcast-fully_connected}
\end{figure}

\emph{Hypercube topology} (figure \ref{fig:topo-hypercube-broadcast-fully_connected}a), along with the Ring and Torus, 
belongs to the group of the most popular migration topologies. The cause of it may be also traced back to the times of 
old parallel systems when slow communication was often a big issue -- as this topology is particularly attractive in 
this context. Namely, it offers the best diameter to number of edges ratio, with the former increasing only 
logarithmically in respect to the number of nodes, at the expense of the number of edges increasing with $O(n\log n)$ 
pace, with very low and homogeneous degree of connectivity -- what eliminates potential bottlenecks (see table 
\ref{tbl:topology-params}). Thus, this topology offers usually the best trade-off between communication delays between 
arbitrary pairs of nodes and the cost represented by the number of edges when the latter is not negligible. This topology is of 
course constructed by putting islands in vertexes of a hypercube of appropriate dimension, and routing connections 
accordingly to the edges of this geometrical structure. This implies, that this topology puts very precise restrictions 
on valid numbers of islands, which must be powers of 2. References to Hypercube topology may be found in \cite{Tanese87, 
Tanese89, Cohoonetal91, CantuPazMejiaOlvera94}. 

Another concept drawn from a context not directly related to the Island Model paradigm is the one behind \emph{Broadcast
topology} (sometimes also called \emph{Star topology}, figure \ref{fig:topo-hypercube-broadcast-fully_connected}b).
This connection layout is rather associated with the Master-Slave model, and thus refers to the fine-grained
parallelization of GAs, as it is the case for instance in \cite{BianchiniBrown93}, but can be occasionally encountered
in non-conventional coarse-grained approaches \cite{Marinetal94}. With very low diameter and linearly growing number of
edges this kind of network may represent a very interesting choice. The biggest problem that is often referred to in the
context of this kind of topology is that the ``center'' node often becomes either communicational or processing
bottleneck.

The limiting case of a migration topology in which all pairs of nodes are directly connected is the \emph{Fully Connected
topology} (figure \ref{fig:topo-hypercube-broadcast-fully_connected}c). While it offers the lowest possible diameter,
the obvious drawback is very quickly increasing number of connections, which may make this choice of topology infeasible
in applications where the numbers of nodes is high. Actually, even though that was not expected, this became also an
issue in our experiments, what forced us to abandon the idea of performing computations for 1024 islands with that
topology. Fully Connected topology is quite often found in the context of Island Model research, with \cite{Petteyetal87,
CantuPazMejiaOlvera94, CantuPazGoldberg00} serving as examples.

\begin{figure}
\centering
	\begin{tabular}{ccc}
		\includegraphics[keepaspectratio=true,width=0.3\linewidth]{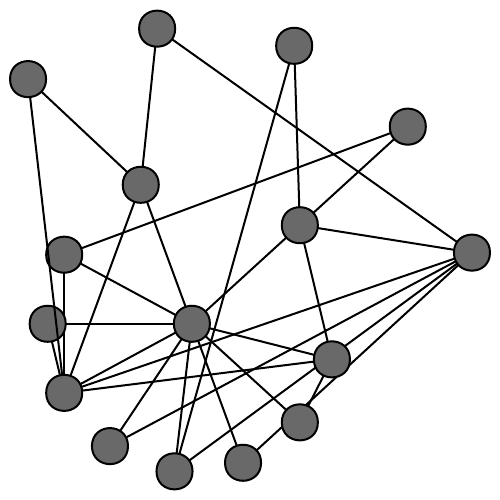}
		&
		\includegraphics[keepaspectratio=true,width=0.3\linewidth]{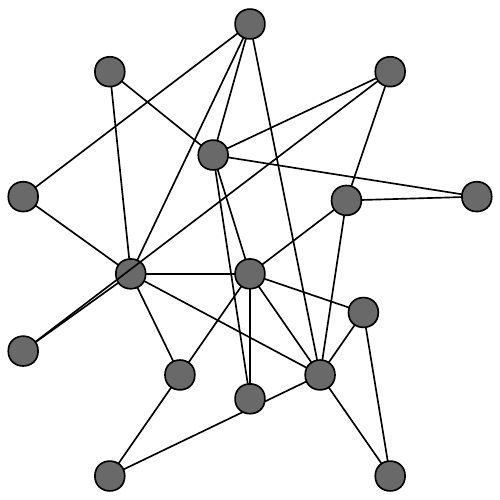}
		&
		\includegraphics[keepaspectratio=true,width=0.3\linewidth]{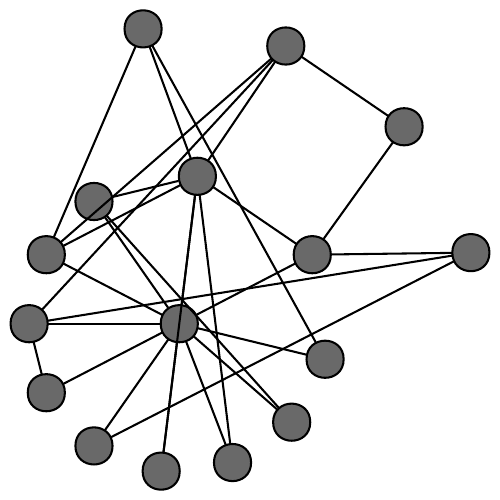}\\
		(a) & (b) & (c)\\
	\end{tabular}
   \caption{Three examples of BA(3,2) topologies with 16 islands.}
\label{fig:topo-ba}
\end{figure}

With this paper we would like to start a trend in Parallel Global Optimization to think 
about the migration topology as of one of fully adjustable algorithm parameters, which choice should not be driven only 
(or mostly) by hardware constraints and historic traditions. To this extent, we included in this study a kind of 
networks that have never been implemented yet in dedicated computing hardware, namely the \emph{scale-free networks}. 
These networks belong to the broader family of \emph{small-world networks}, that is they have small (average) diameter.
In addition to that, in scale-free networks, the number of nodes having a specified degree of connectivity follows
a power law, a property that is found in many natural phenomena (most quoted examples being the network of protein
interactions in cells, the network of hyperlinks in the World Wide Web and the network of sexual contacts between people
\cite{BarabasiAlbert02}). The most known model of scale-free networks creation is the Barab\'asi-Albert model
\cite{BarabasiAlbert99}, which incorporates two features sufficient to obtain a scale-free network: incremental growth
and preferential attachment. We are not going to describe this algorithm here in detail. The only important things in
this context are that the algorithm is not deterministic, and has two parameters: initial cluster size $m_0$ and the
number of links added at one step $m$. For our purposes we used three random seeds to generate three different topologies
for every number of islands using the same values of parameters ($m_0=3$, $m=2$). Example networks with 16 nodes obtained
using the Barab\'asi-Albert algorithm with given parameters for three different random seeds are presented on figure
\ref{fig:topo-ba}. Note that these are shown for illustration purposes only, as such networks are too small to exhibit
statistical properties typical for scale-free networks. To the best of our knowledge, our paper is the first to 
consider scale-free networks in the context of the Island Model. The behavior of Cellular (i.e. fine-grained)
Evolutionary Algorithms with populations structured as Watts-Strogatz and Barab\'asi-Albert networks has been
investigated in \cite{Giacobinietal06}.

Our choice of topologies is certainly not exhaustive, as many other ideas can be found in the related literature. For example
one can often encounter hierarchical approaches that combine a number of topologies as well as different parallelization
paradigms on distinct levels \cite{BianchiniBrown93, CantuPazGoldberg00, KwakJhon07}. The number of combinations that can be 
obtained in this way is however infinite, and thus it has been decided that approaches of this kind are out of the scope of 
this paper. Also we did not consider ideas which are equivalent to an alternative choice of the Island Model parameters.
For example a behavior identical to the one of a \emph{Random topology} \cite{Belding95, Fernandezetal03}, also called Gossip-based in the
literature related to peer-to-peer computing \cite{JelasityBanhelyi09}, can be obtained using a Fully Connected
topology together with the already described ``point-to-point'' migration.

\begin{table}
\centering
\resizebox{0.85\linewidth}{!}{
\begin{tabular}{lccccc}
	\multirow{2}{*}{Topology} & Number   & Valid no. & \multirow{2}{*}{Diameter} & Degree of    & Clustering  \\
	                          & of edges & of nodes  &                           & connectivity & coefficient \\\hline
	Chain           & $n - 1$ & $n \in \mathbb{N}^+$ & $n - 1$ & $\{0,1\}$ & $0$ \\
	One-way Ring    & $n$ & $n \geq 3$ & $n - 1$ & $1$ & $0$ \\
	Ring            & $n$ & $n \geq 3$ & $\big\lfloor n / 2 \big\rfloor$ & $2$ & $0$ \\
	Ring+1+2        & $2n$ & $n \geq 5$ & $\big\lceil \big\lfloor n / 2 \big\rfloor / 2 \big\rceil$ & $4$ & $0.5$ \\
	Ring+1+2+3      & $3n$ & $n \geq 7$ & $\big\lceil \big\lfloor n / 2 \big\rfloor / 3 \big\rceil$ & $6$ & $0.6$ \\
	Cartwheel       & $3n / 2$ & $n = 2k, k \geq 2$ & $\big\lceil n / 4 \big\rceil$ & $3$ & $0$ \\
	Torus           & $3n / 2$ & $n = 2k, k \geq 3$ & $\big\lfloor n / 4 \big\rfloor + 1$ & $3$ & $0$ \\
	Lattice         & $2(n-\sqrt{n})$ & $n=k^2, k \in \mathbb{N}^+$ & $2(\sqrt{n} - 1)$ & $\{2,3,4\}$ & $0$ \\
	Hyperube        & $(n\log_2n)/2$ & $n=2^k, k \in \mathbb{N}^+$ & $\log_2n$ & $\log_2n$ & $0$ \\
	Broadcast       & $n-1$ & $n \in \mathbb{N}^+$ & $2$ & $\{1,n-1\}$ & $0$ \\
	Fully Connected & $n(n-1) / 2$ & $n \in \mathbb{N}^+$ & $1$ & $n - 1$ & $1$\\
	BA(3, 2)        & $2n-3$ & $n \geq 3$ & $\Theta(\ln n / \ln \ln n)^a$ & $[2,n-1]$ & $[0.01, 0.1]^b$\\
\end{tabular}
}
\caption{Parameters of topologies ($n$ is the number of islands). $^a$ See \cite{BollobasRiordan04}. $^b$No theoretical prediction; empirical values for considered numbers of islands are given.}
\label{tbl:topology-params}
\end{table}

\subsubsection{Migration Rate and Frequency}
Separate sets of values of these parameters were used for the two considered base algorithms. These parameters, 
migration frequency especially, are tightly connected to the stopping criterion used, i.e. a limit on the number of 
objective function evaluations performed on one island. This number was determined experimentally based on performance 
of single instances of the base algorithms on considered optimization problems and was fixed to $200\,000$, which proved to be enough for the algorithms to achieve relatively good convergence on these problems. This value remained constant 
regardless of the number of islands in the parallel algorithm, which allowed keeping the migration frequency on the same 
level for all considered numbers of islands. The migration frequency itself was determined based on the amount of work 
performed by each base algorithm per unit of time. For the DE algorithm, the migration was set to occur every 100 
generations, i.e. 2000 objective function evaluations for the assumed population size of 20. For the SA algorithm, 
the migration took place at the end of each annealing cycle ($10\,000$ algorithm iterations). Migration rate was fixed 
to 10\% of the population in case of DE (i.e. 2 individuals) and to 1 individual for SA (which is of course the only 
possible choice, as this algorithm is not population-based). 

\subsubsection{Migration Algorithm}
We used the asynchronous migration as we aim to consider large numbers of islands -- this kind of migration is much 
more likely to be implemented in a large distributed environment than the synchronized one. We decided to reflect 
in the migration the philosophy of elitism -- from every island the best individuals were selected for 
migration, and the incoming individuals replaced the worst ones in the population provided that they were better. Simulated
Annealing was treated as an algorithm with population of size 1 (the current solution was sent to the neighbor islands
and it was replaced by an incoming immigrant if and only if the latter had better fitness value).
Finally, (copies of) migrating individuals were sent to all neighbors of the source island (which is a ``broadcasting'' 
migration scheme) instead of the one randomly selected (``point-to-point'' migration). It may be worth noting that during
preliminary experiments conducted within this study, the ``point-to-point'' migration model was applied. Such a choice was
a result of the fact that this kind of communication was used in one of the previous studies conducted by our team, which
was focused on Global Optimization in peer-to-peer computing environments \cite{JelasityBanhelyi09}. We then opted for
``broadcasting'' migration as the latter should emphasize the impact of the migration topology on the behavior of the
parallel algorithm. This can be deduced from an observation that the more neighbors an island has, the more rarely
it will communicate with one particular neighbor in the former model, and thus noticing differences caused by changed
migration topology would require much more migration periods than when the latter approach is used. This prediction has
been confirmed by obtained experimental results.

\subsection{Analysis and Interpretation of Results}
\label{sec:methodology-results-analysis}

Summarizing information given in previous subsections, the computational experiments presented in this study included 
testing Parallel Global Optimization Algorithms for 14 topologies, 4 distinct numbers of islands, 3 test problems and 3 
base algorithms (counting the SA algorithm with untuned and tuned parameters as separate) which constitutes the total 
of 36 distinct set-ups in which each one of the 14 topologies was tested. Calculations for every set-up were repeated 30 
times in order to allow for statistical analysis. 

The main output from the computational experiments was the mean and standard deviation of the objective function value 
of the final solution obtained in each set-up in the aforementioned 30 repetitions. Obtained differences between final 
objective function values were tested for statistical significance using Welch's t-test\footnote{The number of trials 
performed (30) is sufficient to treat tested variables as normally distributed, but it cannot be assumed that they all 
have equal variances, thus the use of the Welch's modification in place of the traditional t-test.} \cite{Welch47} at 
confidence level $0.01$ and used to introduce a preorder over the set of topologies for each set-up. These preorders are 
``partial rankings'' of migration topologies by their performance for every test problem, base algorithm and number of 
islands. They are not rankings in the mathematical sense, because statistically significant differences may not appear
in all cases, leaving in each set-up certain number of pairs of topologies incomparable\footnote{It may be important to stress that 
the lack of statistically significant difference between two results does not mean they are identical. It just means 
that the difference is too small to be detected in the performed tests and thus there is no indication to rank one 
result over another. This should not be interpreted as assuming equality of the results. The difference may show up when 
the larger sample size is used to perform the test. On the other hand, from an application point of view, it is worth to 
consider topologies as different only if their difference shows up in the statistics of small samples.}.

Having preorders of topologies for every problem-base algorithm-number of islands triple, several interesting questions 
may be posed. For example, preorders may be grouped by every pair of these three variables in order to see how do they
change along with the remaining third parameter (e.g does the number of islands affect significantly the ranking of 
topologies/the optimal choice of topology). In order to answer these questions one needs a mathematically sound method 
of preorders comparison. For this purpose the Kendall $\tau$ rank correlation coefficient \cite{Kruskal58} has been 
used in this paper. This non-parametric statistic allows measuring the degree of correspondence between two rankings; it 
takes values from the$[-1,1]$ interval, with $1$ meaning that compared rankings are identical, $-1$ meaning the rankings 
are exactly opposite, and $0$ meaning that the rankings are unrelated. As in this paper we compare preorders instead of 
rankings, the $\tau_b$ variant of the statistic, suitable for such situation, has been used. 

It may be worth noting that the value of Kendall $\tau_b$ depends not only on the number of elements ranked in the same 
way in compared preorders, but also on the number of ties in each of them. Thus, value of the Kendall $\tau_b$ lower 
than $1$ may not only indicate that some elements are ranked differently, but also that compared preorders differ with 
the ``degree of completeness'', specificity. This property of the Kendall $\tau_b$ is a reason for which, despite being a 
very useful and elegant tool, its value alone is not sufficient to determine what is the exact \emph{nature} of the 
difference between compared preorders. In order to find it out, either visual inspection of graphs equivalent to these
preorders (e.g. using the Hasse diagrams) or direct analysis of numbers of discordant pairs and ties in compared
preorders is necessary.

Finally, even if two preorders are not identical, they can still give exactly the same clue about which are the best
and the worst elements in the preordered set. Thus, when one is interested in a subset of elements that is ranked consistently
within multiple preorders, one has to perform even more detailed analysis, looking at exact positions of
particular elements in each of these preorders.

Such analysis of preorders is of course the more meaningful the more specific the preorders are -- the less number of 
incomparable elements they contain\footnote{Alternatively one can say about the \emph{height} of a preorder -- defined 
as the length of the longest path in a Direct Acyclic Graph (DAG) equivalent to that preorder. Such graph is possible to 
be created for any preorder. Height of a preorder says what is the largest number of elements from the preordered set 
that can be put in a sequence so that every element of this sequence is ranked higher than all its successors.}. 
Unfortunately during the preliminary analysis of the results of computational experiments described in this paper it 
turned out that the basic criterion of results comparison described in the beginning of this section (i.e. the one based 
only on the mean and standard deviation of the final objective function value) turned out to be not discriminating 
enough in some cases -- certain obtained preorders contained many ties, and their height was very low (2-3). This 
occurred in situations when in a particular set-up for most of the migration topologies the algorithm converged 
sufficiently close to the global optimum of the objective function. In order to deal with this problem, an extended 
criterion of algorithm comparison has been introduced, based on the average algorithm performance at the end of every 
migration period. When the average final solutions of two algorithms were not distinguishable, comparison between them 
was based on earlier stages of optimization (namely on the latest one in which the difference was statistically 
significant). This approach is compatible with the basic one (the last migration period is at the same time the end of 
processing), i.e. a preorder obtained using the extended method will always contain the preorder constructed using the 
basic method -- and thus preorders obtained with these two methods can be meaningfully compared.

\begin{figure}
\centering
	\begin{tabular}{cc}
		\includegraphics[keepaspectratio=true,width=0.45\linewidth]{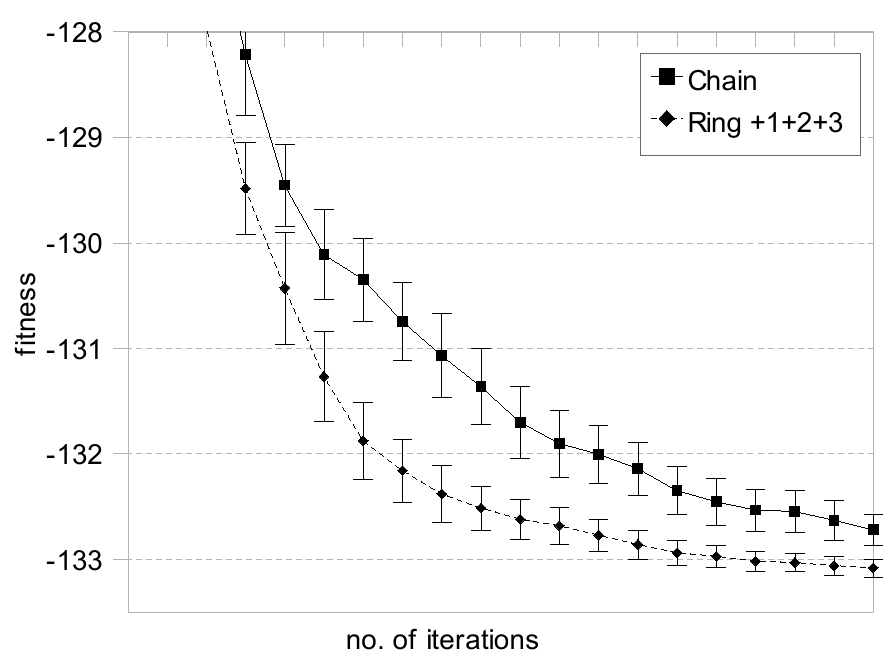}
		&
		\includegraphics[keepaspectratio=true,width=0.45\linewidth]{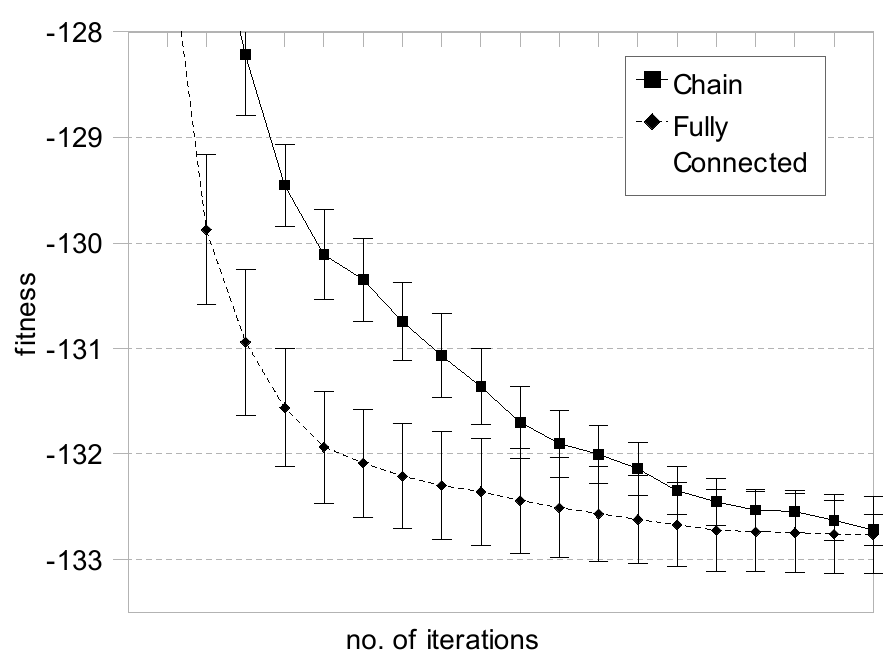}\\
		(a) & (b)\\
	\end{tabular}
   \caption{Illustration of a situation where the extended criterion of the results comparison was used. Sometimes the
   final value of the objective function is sufficient to discern between results (a), but in the other cases the earlier
   stages of the optimization have to be taken into account (b).}
\label{fig:evolog-examples}
\end{figure}

The idea of the extended method of results comparison is illustrated on figure \ref{fig:evolog-examples}.
The data are taken from the actual optimization process for the $V_{31}$, $SA_t$, 128 islands set-up. The plots show the 
fitness value of the best individual over all islands after each migration period averaged over 30 runs, together with 
standard deviation bars. Figure \ref{fig:evolog-examples}a shows a typical situation where the difference in the final 
value of the objective function is large enough to provide statistical evidence of the difference between algorithms. In 
situations where this is not the case, as shown on figure \ref{fig:evolog-examples}b, considering earlier stages of 
optimization often provides clues that even despite similar final results, one algorithm still should be ranked over 
another because it reaches the final solution faster.

This extended method of algorithms performance assessment may however produce relations that are not preorders. Because 
the comparison between different pairs of algorithms may be based on their states at different points of time, the 
constructed relation may violate transitivity as well as contain cycles. analysis of the gathered data revealed that such 
anomalies appear exclusively when statistical tests pass barely beneath a-priori assumed confidence level and/or when 
comparisons are based on very early stages of optimization\footnote{The latter should not be a surprise, because early 
stages of optimization are much more strongly affected by random initial conditions than the final ones, and sample 
count sufficient for reliable analysis of the final results of the algorithm may not be sufficient to perform reliable 
tests of its average performance at the early stages due to relatively high variance of the data.}. Thus, in order to 
deal with the problem of invalid preorders, one can both adjust the confidence level of the statistical tests to prune 
out uncertain situations and limit the data available for algorithm comparison to the more reliable final stages. Both 
these approaches will result in more conservative ordering of elements, with incomparability appearing in place of 
previous questionable situations.

\subsection{Computing Platform}
The experiments were run on a software platform called PaGMO \cite{PaGMO09}, developed within the Advanced Concepts 
Team at the European Space Agency and initially conceived in the context of Global Optimization of spacecraft 
trajectories \cite{Izzoetal07,Vinkoetal07,Yametal09}. Written in C++, PaGMO implements the Island Model framework 
through the use of multiple threads of execution and communication via the shared memory mechanism, so that optimization
on islands map directly to Operating System threads. Through the use of object-oriented programming paradigms, such as
inheritance and dynamic polymorphism, a hierarchy of objects is defined as follows: 
\begin{itemize}
\item a problem is defined by a function accepting a vector of values as input (i.e., a chromosome) and returning a double-precision value (i.e.,
the fitness) as output;
\item a population is defined as a collection of chromosomes coupled with a problem;
\item an island is defined by a population coupled with an algorithm used for the optimization of the chromosomes in the problem associated to the population;
\item an archipelago is a collection of islands coupled with
\begin{itemize}
\item a topology object, describing how the islands are connected to each other,
\item a migration scheme object, specifying the details of the migration algorithm.
\end{itemize}
\end{itemize}
The evolution of each island in an archipelago proceeds asynchronously with respect to the other islands of the 
archipelago, the only synchronization point (implemented through mutexes) being the exchange of information between 
islands during migration. 

In order to ensure maximum portability, PaGMO relies heavily on the facilities provided by the Boost libraries 
\cite{boost09}, in particular regarding random number generators and threading libraries. It is hence possible to 
compile and run PaGMO both on Unix-like and Windows systems. The experiments presented in this paper were run on an 
8-core Apple OSX 64-bit server equipped with Intel Xeon processors, using the GCC compiler version 4.0.1. PaGMO is Free 
Software, released under the terms of the GNU public license. 

\section{Results Summary}
\label{sec:results} 

In this section we present the summary of obtained results and drawn conclusions. The outline of this section
resembles the one of the section \ref{sec:methodology-results-analysis}, in which the methodology of results analysis
has been described.

\begin{figure}
\centering
   \includegraphics[keepaspectratio=true,width=\linewidth]{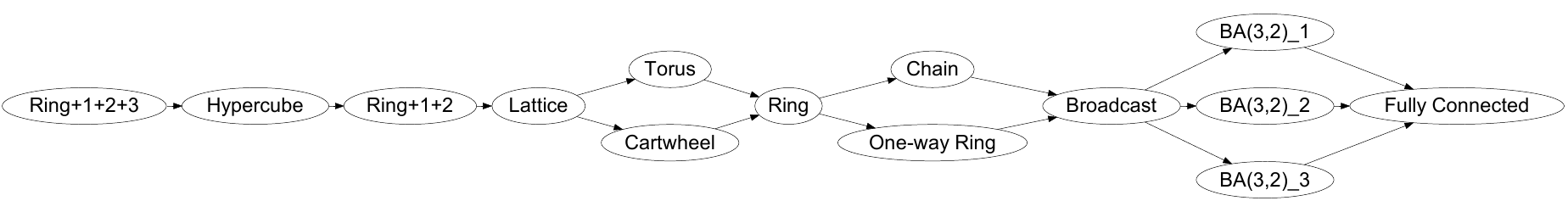}
   \caption{Example of a preorder, showing statistically significant differences in performance of
   parallel DE algorithms with 512 islands which use different migration topologies, obtained for the Schwefel function
   minimization problem.}
\label{fig:sample-ranking}
\end{figure}

As it was presented in the aforementioned section, parallel algorithms using different migration topologies were ranked
accordingly to their performance in 36 distinct set-ups, being all combinations of 3 considered parameters: the number
of islands, the optimization problem and the base algorithm. Because of the low quality of rankings constructed using
the basic method of algorithm comparison, in 20 cases the decision was taken to rank topologies accordingly to the 
extended criterion. However, obtaining valid rankings required adjusting the confidence level of statistical tests
and the amount of data used for comparison. Final parameters of the extended procedure of preorder construction are
given in table \ref{tbl:preorder-construction-params}.
\begin{table}
\centering
\begin{tabular}{cccccc}
No. of & $f_R$ & $f_S$ & \multicolumn{3}{c}{$V_{31}$} \\
islands & \scriptsize{$SA_t$} & \scriptsize{$SA_t$} & \scriptsize{$DE$} & \scriptsize{$SA_u$} & \scriptsize{$SA_t$}\\\hline
128  & $0.001$ & $0.01$ & $0.01^a$ & $0.01^a$ & $0.01$ \\
256  & $0.01$ & $0.01$ & $0.01^a$ & $0.01^a$ & $0.001$ \\
512  & $0.01$ & $0.01$ & $0.01$ & $0.001^a$ & $0.0001$ \\
1024 & $0.01$ & $0.01$ & $0.01$ & $0.01^a$ & $0.001$ \\
\end{tabular}
\caption{Confidence level used in statistical tests in the extended procedure of preorder construction. In set-ups not
included in the table the basic procedure was used. $^a$Only second half of the optimization process was considered.}
\label{tbl:preorder-construction-params}
\end{table}
Example preorder is presented on figure \ref{fig:sample-ranking}.

Constructed preorders were compared using Kendall $\tau_b$ coefficient which value tells if the preorders are correlated
positively, negatively or are unrelated. First the coefficient was evaluated for preorders obtained from the same optimization
problem and base optimization algorithm, in order to assess if changing the number of islands involved in the 
optimization causes strong perturbations in the way migration topologies are ordered. The results of this comparison are
reported in table \ref{tbl:kendalltau-islands}.
\begin{table}
\centering
\resizebox{0.75\linewidth}{!}{
\begin{tabular}{ccccccccccc}
& & \multicolumn{3}{c}{$f_R$} & \multicolumn{3}{c}{$f_S$} & \multicolumn{3}{c}{$V_{31}$} \\
                        & & \scriptsize{256} & \scriptsize{512} & \scriptsize{1024} & \scriptsize{256} & \scriptsize{512} & \scriptsize{1024} & \scriptsize{256} & \scriptsize{512} & \scriptsize{1024}\\
\multirow{3}{*}{$DE$}   & \scriptsize{128} & $0.88$ & $\textbf{0.75}$ & $\textbf{0.73}$ & $0.93$ & $0.89$ & $0.86$ & $0.91$ & $0.82$ & $0.8$ \\
                        & \scriptsize{256} &  & $0.94$ & $0.90$ &  & $0.95$ & $0.92$ &  & $0.89$ & $0.88$ \\
                        & \scriptsize{512} &  &  & $0.96$ &  &  & $0.88$ &  &  & $0.99$ \\
\multirow{3}{*}{$SA_u$} & \scriptsize{128} & $0.94$ & $0.92$ & $0.92$ & $0.98$ & $0.97$ & $0.93$ & $\textbf{0.77}$ & $0.88$ & $0.82$ \\
                        & \scriptsize{256} &  & $0.95$ & $0.93$ &  & $0.99$ & $0.95$ &  & $0.84$ & $\textbf{0.79}$ \\
                        & \scriptsize{512} &  &  & $0.94$ &  &  & $0.98$ &  &  & $0.83$ \\
\multirow{3}{*}{$SA_t$} & \scriptsize{128} & $0.89$ & $0.9$ & $0.84$ & $0.96$ & $0.89$ & $0.88$ & $0.82$ & $0.85$ & $0.81$ \\
                        & \scriptsize{256} &  & $0.92$ & $0.84$ &  & $0.92$ & $0.91$ &  & $0.86$ & $\textbf{0.78}$ \\
                        & \scriptsize{512} &  &  & $0.84$ &  &  & $1.0$ &  &  & $\textbf{0.76}$ \\
\end{tabular}
}
\caption{Kendall $\tau_b$ values for rankings compared for different numbers of islands. Values lower than $0.8$ are highlighted with bold.}
\label{tbl:kendalltau-islands}
\end{table}
Computed Kendall $\tau_b$ values varied from $0.73$ to $1$ with median $0.89$. Only in six (out of 54) cases the value 
of the statistic was lower than $0.8$. In one situation, topologies were ordered exactly in the same way for two 
different numbers of islands ($f_S$, $SA_t$, $512$ and $1024$ islands). This indicates that ordering of topologies 
remains very consistent when only the number of islands changes and the other parameters remain the same. In other 
words, the number of islands in the parallel algorithm has a vary limited impact on the ordering of topologies, and thus 
definitely is not a decisive factor when it comes to the choice of the optimal migration topology in a particular 
situation. More about the nature of the differences that take place when the number of islands is changed is revealed 
when one analyzes the numbers of discordant pairs in compared preorders. In 70\% of the cases (38 out of 54) there 
were no discordant pairs, and the maximum encountered number was 7 (with maximum possible being 91). This suggests that 
when the number of islands changes, the specificity of the ranking is altered rather than the actual ordering of 
topologies. Further analisis revealed that in more than 80\% of the cases (44 out of 54) the ordering of topologies 
created for a bigger number of islands contained less ties than the one obtained for a lower one. The conclusion is that 
in the analyzed set-ups increasing the number of islands in general leads to obtaining more detailed ranking. This 
confirms the intuition presented in section \ref{sec:methodology-no-of-islands} -- that the greater the number of 
islands is, the more accented the differences between topologies are. This in turn means, that the more islands are to 
be used, the more important the choice of the migration topology will be. 

A similar analysis was performed to assess the ordering sensitivity to the optimization problem. Results of the comparison of
preorders are reported in table \ref{tbl:kendalltau-problem}.
\begin{table}
\centering
\resizebox{0.75\linewidth}{!}{
\begin{tabular}{cccccccccc}
& & \multicolumn{2}{c}{128} & \multicolumn{2}{c}{256} & \multicolumn{2}{c}{512} & \multicolumn{2}{c}{1024} \\
& & \scriptsize{$f_S$} & \scriptsize{$V_{31}$} & \scriptsize{$f_S$} & \scriptsize{$V_{31}$} & \scriptsize{$f_S$} & \scriptsize{$V_{31}$} & \scriptsize{$f_S$} & \scriptsize{$V_{31}$} \\
\multirow{2}{*}{$DE$}            & \scriptsize{$f_R$} & $\textbf{0.65}$ & $0.84$ & $0.82$ & $0.91$ & $0.81$ & $0.95$ & $0.86$ & $0.97$ \\
                               & \scriptsize{$f_S$} &  & $0.84$ &  & $0.84$ &  & $0.76$ &  & $0.8$ \\
\multirow{2}{*}{$SA_u$} & \scriptsize{$f_R$} & $0.8$ & $\textbf{0.21}$ & $0.86$ & $\textbf{0.22}$ & $0.84$ & $\textbf{0.15}$ & $0.92$ & $\textbf{0.6}$ \\
                               & \scriptsize{$f_S$} &  & $\textbf{0.39}$ &  & $\textbf{0.33}$ &  & $\textbf{0.25}$ &  & $\textbf{0.68}$ \\
\multirow{2}{*}{$SA_t$}   & \scriptsize{$f_R$} & $0.89$ & $0.83$ & $0.93$ & $0.83$ & $0.9$ & $0.76$ & $0.82$ & $0.76$ \\
                               & \scriptsize{$f_S$} &  & $0.78$ &  & $0.81$ &  & $\textbf{0.65}$ &  & $0.82$ \\
\end{tabular}
}
\caption{Kendall $\tau_b$ values for rankings compared for different optimization problems. Values lower than $0.75$ are
highlighted with bold.}
\label{tbl:kendalltau-problem}
\end{table}
The Kendall $\tau_b$ values varied from $0.15$ to $0.97$ with median $0.81$. Larger range of variation
together with lower median indicate that the influence of the optimization problem on the way topologies are ranked
is stronger than in the case of the number of islands parameter. Remarkably though, in more than a half of the cases, the
correlation between obtained rankings was very strong, and in all cases calculated coefficient was positive -- only in 6
situations (out of 36) orderings could be called slightly correlated or unrelated. As can be seen from the distribution
of the lowest values in table \ref{tbl:kendalltau-problem}, least related rankings occurred mostly when SA algorithm with
untuned parameters was used, and almost exclusively in situations where one of the compared rankings was obtained for
the Lennard-Jones potential minimization problem. analysis of the discordant pairs and numbers of ties in preorders do
not provide as clear clues as in the case of the number of islands. The only two evident things are the
following. First, for the $SA_u$ algorithm (i.e. where the low values of the Kendall $\tau_b$ coefficient are
concentrated) preorders obtained for the $V_{31}$ problem were always much less specific than those obtained for other
problems, indicated by the numbers of ties. Second, in the same cases, compared preorders contained significantly more
discordant pairs than in situations corresponding to the rest of cells in the table. Possible reasons of this situation
are very low differences between final objective function values obtained with the $SA_u$ algorithm for the $V_{31}$
problem for different topologies, together with their relatively high standard deviation -- which resulted in rather
vague preorders. Taking into account general results of this comparison, the impact of the considered optimization
problems on the shape of the rankings of topologies is stronger than in the case of the number of islands, but still far
from dramatic.

Finally, the impact of the optimization algorithm on the topology ranking was assessed using the same methodology.
Calculated Kendall $\tau_b$ values are reported in table \ref{tbl:kendalltau-algorithm}.
\begin{table}
\centering
\resizebox{0.75\linewidth}{!}{
\begin{tabular}{cccccccccc}
& & \multicolumn{2}{c}{128} & \multicolumn{2}{c}{256} & \multicolumn{2}{c}{512} & \multicolumn{2}{c}{1024} \\
& & \scriptsize{$SA_u$} & \scriptsize{$SA_t$} & \scriptsize{$SA_u$} & \scriptsize{$SA_t$} & \scriptsize{$SA_u$} & \scriptsize{$SA_t$} & \scriptsize{$SA_u$} & \scriptsize{$SA_t$}\\
\multirow{2}{*}{$f_R$}            & \scriptsize{$DE$} & $\textbf{-0.53}$ & $\textbf{-0.35}$ & $\textbf{-0.48}$ & $\textbf{-0.31}$ & $\textbf{-0.44}$ & $\textbf{-0.34}$ & $\textbf{-0.28}$ & $\textbf{-0.25}$ \\
                               & \scriptsize{$SA_u$} & & $0.83$ & & $0.86$ & & $0.88$ & & $0.87$ \\
\multirow{2}{*}{$f_S$} & \scriptsize{$DE$} & $\textbf{-0.2}$ & $\textbf{-0.24}$ & $\textbf{-0.19}$ & $\textbf{-0.23}$ & $\textbf{-0.13}$ & $\textbf{-0.09}$ & $0$ & $0.09$ \\
                               & \scriptsize{$SA_u$} & & $0.98$ & & $0.96$ & & $0.91$ & & $0.92$ \\
\multirow{2}{*}{$V_{31}$}   & \scriptsize{$DE$} & $0.34$ & $\textbf{-0.18}$ & $0.27$ & $\textbf{-0.32}$ & $0.31$ & $\textbf{-0.17}$ & $0.19$ & $\textbf{-0.02}$ \\
                               & \scriptsize{$SA_u$} & & $0.49$ & & $0.44$ & & $0.4$ & & $0.83$ \\
\end{tabular}
}
\caption{Kendall $\tau_b$ values for rankings compared for different base optimization algorithms. Negative values are
highlighted with bold.}
\label{tbl:kendalltau-algorithm}
\end{table}
This time the coefficient varied from $-0.53$ to $0.98$ with median $-0.01$. Negative as well as very close to zero 
values of the coefficient appear regularly for the first time. Closer look at the table \ref{tbl:kendalltau-algorithm} 
reveals that these values were found exclusively in comparisons between the $DE$ algorithm with one of the versions of 
the $SA$ algorithm, while high values show up only when comparing rankings obtained with the $SA$ algorithm with tuned 
and untuned parameters. Motivated by this fact, more detailed analysis has been done for these two cases separately. In 
the comparisons between the two variants of the SA algorithm used, Kendall $\tau_b$ varied from $0.4$ to $0.98$ with 
median $0.87$, showing that tuning the parameters of the SA algorithm has a little impact on the topology ranking, as 
preorders stay strongly correlated in most cases, sometimes dropping to slight positive correlation (3 lowest numbers of 
islands for the $V_{31}$ problem). In turn, when fundamentally different base algorithms were compared (DE against SAs), 
values of the coefficient varied from $-0.53$ to $0.34$ with median $-0.19$, what indicates something between slight 
negative correlation and the lack of correlation. Not surprisingly, in the latter comparison there are much more 
discordant pairs between rankings than in the former. For DE-SA comparison, up to 51 such pairs were observed with 
median around 40, while the maximum of discordant pairs between two variants of SA was 11 with very low median of 1. 
Clearly the differences between rankings of topologies observed here were highest in the whole experiment. This shows 
that the change of the base algorithm altered topology rankings much more than the other two parameters. Moreover, the 
amount of the changes in the preorder depends on the character of the change of the algorithm -- simple parameter tuning 
did not perturb rankings strongly (at the level comparable with the other parameters), but when the algorithm was 
changed completely, this resulted in fundamentally different ordering of migration topologies. The conclusion is that if 
an optimal migration topology for computation is to be chosen, the first factor that has to be taken into account is the 
base algorithm. 

\begin{figure}
\centering
	\begin{tabular}{cc}
		\includegraphics[keepaspectratio=true,width=0.375\linewidth]{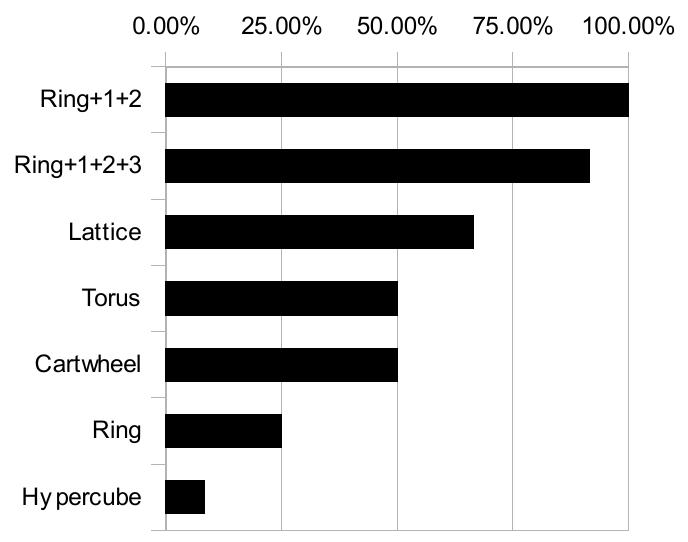}
		&
		\includegraphics[keepaspectratio=true,width=0.375\linewidth]{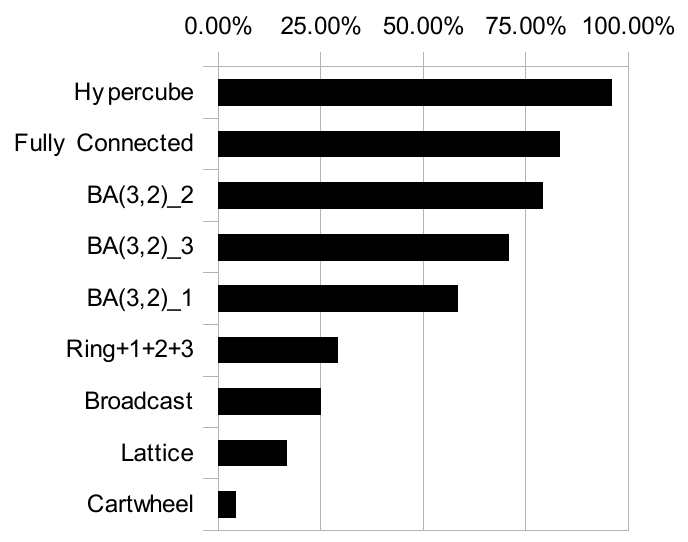}\\
		(a) & (b)\\
	\end{tabular}
   \caption{Best migration topologies accordingly to rankings obtained for DE (a) and SA (b).}
\label{fig:best-topologies}
\end{figure}

\begin{figure}
\centering
	\begin{tabular}{cc}
		\includegraphics[keepaspectratio=true,width=0.375\linewidth]{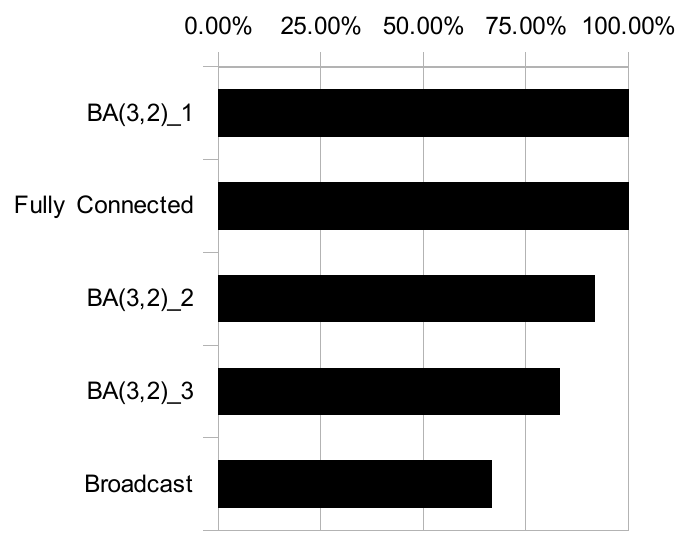}
		&
		\includegraphics[keepaspectratio=true,width=0.375\linewidth]{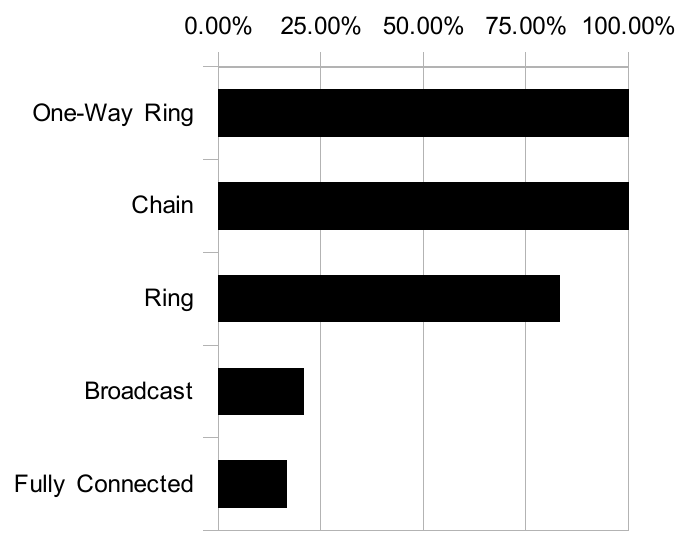}\\
		(a) & (b)\\
	\end{tabular}
   \caption{Worst migration topologies accordingly to rankings obtained for DE (a) and SA (b).}
\label{fig:worst-topologies}
\end{figure}

A further analysis of the preorders is now presented aiming at extracting from the gathered data a clear clue on good 
topology choices for the two considered base algorithms. This has been done by extracting 3 best and 3 worst topologies 
from every preorder\footnote{``Best'' and ``worst'' topologies in a preorder are defined as those which are 
significantly better (worse) than the greatest number of other topologies. In consequence, a topology which is 
incomparable with all others will be neither considered good nor bad.} and, for both algorithms, counting how often each 
topology appears in ``top'' and ``bottom'' 3. Results of the analysis are presented on figures \ref{fig:best-topologies} 
and \ref{fig:worst-topologies}. As seen from the figures, for both DE and SA it is easier to answer which topologies 
\emph{not} to use with these algorithms -- the numbers of topologies which appear within the bottom 3 of the rankings is 
smaller than of those showing at the top. 


In case of the DE algorithm, the safest choice of the migration topology seems to be one of the extended Rings: +1+2+3 
or +1+2. The topology ranked as third, Lattice, shows up in the top 3 already in less than three quarters of set-ups. 
Clearly one should avoid using with DE algorithm topologies which are characterized by lowest diameters. This seems to 
indicate that parallel DE algorithm does not benefit from a quick spread of information between islands and that 
evolutionary niches need to be kept uncontaminated by the best genes in order to allow the correct balance between 
exploitation and exploration. Interestingly, for SA the situation is the contrary (fact that can be also inferred 
from the negative Kendall $\tau_b$ values obtained when comparing the preorders for the base algorithms 
\ref{tbl:kendalltau-algorithm}). The Hypercube topology is an unquestionable winner for this algorithm and all BA 
networks appear in the top 3 of the rankings quite frequently. Fully Connected topology looks as another safe choice, 
but only at the first glance -- it also appears in the set of the worst topologies. Parallel SA algorithm evidently does 
not go along well with topologies of high diameter (One-way Ring and Chain always in the worst 3), and, contrary to 
parallel DE, seems to work better when the information is spread relatively fast between islands. With this respect the 
Barabasi-Albert networks offer a good compromise between information spread and number of links (this last being related 
to the communication overhead introduced by the migration operator). 

Finally, an attempt was made to find out if some of the migration topology parameters reported in table 
\ref{tbl:topology-params} correlate with results obtained for the two considered algorithms. Such analysis could provide 
insight about what properties make some topologies work better than others for the given algorithm. For this purpose we 
ranked all topologies accordingly to the following parameters: graph density (or, equivalently, the number of edges), 
average shortest path length between nodes, graph diameter, average and maximum node degree, and the clustering 
coefficient \cite{WattsStrogatz98}. This has been done for each number of islands separately, because non-deterministic 
character of the Barabasi-Albert topologies caused changes in rankings with the change of the number of islands; the 
changes were however marginal, amounting to maximum 1 discordant pair or 2 ties depending on the situation -- only 
relative positions of the three scale-free topologies were affected. 

Two of the chosen graph parameters, graph density and average node degree, turned out to be equivalent in terms of the
order they introduce among considered migration topologies. Preorders obtained using average shortest path length,
graph diameter, and maximum node degree were also strongly correlated with each other (absolute values of Kendall
$\tau_b \geq 0.91$), with the latter producing preorders opposite to the former two (which is not a surprise when one
takes into account definitions of these measures). In the remaining situations obtained Kendall $\tau_b$ values
oscillated around the area of slight correlation ($0.5$) rather than that of the lack of correlation ($0$). This might
indicate that the set of selected graph features could be extended\footnote{Of course in an ideal situation all selected
features would be uncorrelated.}.

\begin{figure}
\centering
	\begin{tabular}{cc}
		\includegraphics[keepaspectratio=true,width=0.43\linewidth]{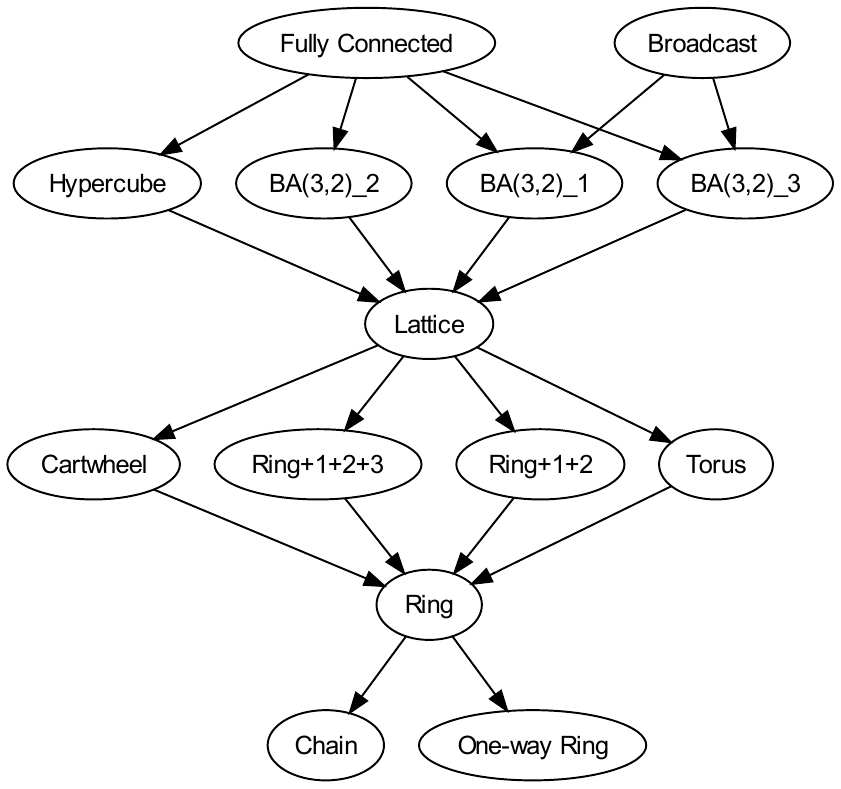}
		&
		\includegraphics[keepaspectratio=true,width=0.29\linewidth]{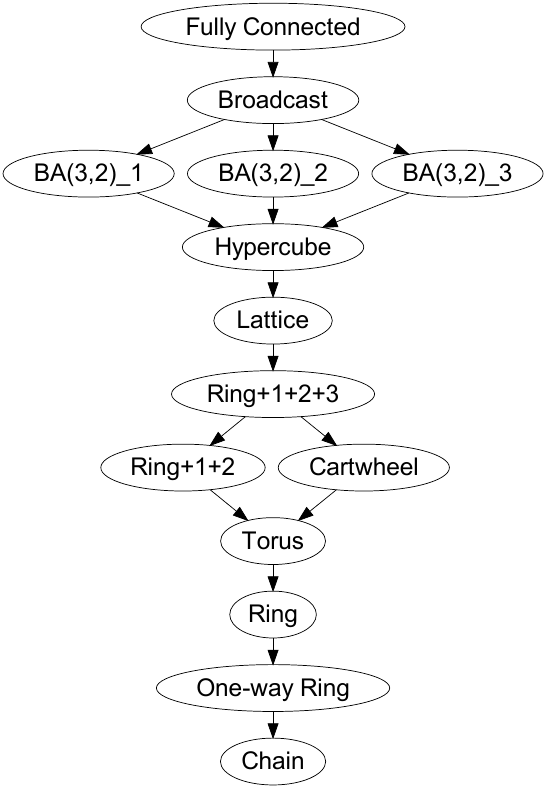}\\
		(a) & (b)\\
	\end{tabular}
	\caption{Topologies ranked by performance in the $f_R$, $SA_u$, 512 islands set-up (a) and topologies ordered by the
		ascending network diameter, also 512 islands (b).}
\label{fig:asa-vs-diameter}
\end{figure}

Comparison of preorders of topologies by their features with those obtained as results of conducted computational
experiment yielded interesting results, especially in the case of the parallel SA algorithm. Strongest correlation with
preorders of topologies by performance in parallel SA was discovered for average shortest path length, graph diameter,
and maximum node degree measures (negative for the two former, positive for the latter). The absolute values of Kendall
$\tau_b$ varied from around $0.6$ to $0.9$ with median $0.77$ (with small differences for each of the three features),
when preorders obtained for $SA_u$, $V_{31}$ set-ups were not taken into account\footnote{For this particular situation
values of the coefficient were much lower, ranging from around $0.09$ to $0.4$. This is not the first time that data
obtained in this particular set-up stand out from the rest -- as it was already mentioned, the reason for this may be
the low quality of obtained preorders (low height, many ties), which in turn was most probably effected by the usage of
untuned algorithm parameters for a fairly difficult optimization problem.}. Certain amount of correlation was also found
for the graph density feature (or equivalently to the average node degree). While the values of the Kendall $\tau_b$
coefficient were smaller than in the previous case, they were more consistent over the set of all obtained empirical
preorders, as (including the problematic $SA_u$, $V_{31}$ set-up) they ranged from $0.39$ to $0.86$ with median $0.69$.
At any rate, all these results confirm earlier observation that in conducted experiments the parallel SA algorithm
performed better on more tightly connected topologies. Two preorders for which the strongest correlation has been found
during this analysis are presented on figure \ref{fig:asa-vs-diameter}.
Kendall $\tau_b$ coefficient for these preorders equals to $0.93$, and they contain no discordant pairs, what means that
they order topologies in an exactly the same way, differing only with specificity.

In case of the parallel DE algorithm none of the analyzed measures correlated with algorithm performance enough to
unambiguously identify one feature which turns out to be the most important. Numerically, the highest values of
Kendall $\tau_b$ were found for the same triple of features as in the case of the SA algorithm (average shortest path
length, graph diameter, and maximum node degree), but with an opposite sign and magnitude ranging from $0.03$ to $0.6$
with median $0.36$. While the opposite sign confirms observations done in the previous parts of this section, the
magnitude does not allow stating that in the experiments parallel DE's preference of topologies was exactly opposite to
the one of parallel SA. Because discovered correlation is at the best slight (in more than half of the cases there's
definitely none), certainly some other factors, not discovered by the considered set of migration topology features,
came into play.

\section{Conclusions and Future Prospects}
\label{sec:conclusions-and-future}

Our experiments show systematically that the migration topology is an important parameter of a Parallel Global 
Optimization Algorithm which uses the Island Model. The migration topology affects the quality of the obtained final 
solution and the pace of convergence. The impact of the migration topology on the overall optimization process appears 
evident in large networks such the ones considered in this work and will thus play an increasingly important role for 
algorithms deployed on future multi-core architectures or in distributed computing projects. 

We built migration topology rankings (or, more precisely, preorders) for different numbers of islands, different 
optimization problems and different base algorithms, the latter having unquestionably the biggest impact on the 
resulting rankings. While our conclusions are limited to the algorithms and problems considered in the paper, obtained 
preorders have a great practical value. For the two tested algorithms, it was possible to identify limited subsets of 
topologies that performed best and worst over the set of considered optimization problems allowing to conclude that 
certain kinds of topologies are to be avoided in both cases, while certain others are good candidates as a first choice. 
Parallel Differential Evolution seems to prefer the Ring+1+2 and the Ring+1+2+3 topologies over others and to dislike 
topologies such as Fully Connected or Barabasi-Albert networks where a fast information spread over the entire network 
is allowed. For Parallel Simulated Annealing with Adaptive Neighborhood a good topology choice is the Hypercube or 
Barabasi-Albert networks, while the One-way Ring, Chain and Ring topologies are to be avoided. The result on the 
Barabasi-Albert topologies is of particular interest, because -- contrary to purely regular topologies which have to be 
constructed (and maintained, in case of a distributed system) algorithmically -- this kind of networks is expected to 
\emph{emerge} in a dynamic, growing system, and thus may represent an attractive option from the point of view of 
peer-to-peer or grid computing.

Future work includes performing similar experiments and analysis for a broader set of optimization problems,
including some real engineering optimization cases on top of extended sets of benchmark functions. The research can also
be extended to cover more base algorithms. Particularly interesting could be an answer to the question if different
variants of the same algorithm prefer different migration topologies and if different algorithms can be placed on top
of the same network profiting of its topology. Such findings on any particular algorithm provide valuable clues for
performing theoretical analysis of the behavior of its parallel version, similar to the one done by Cant\'u-Paz for
parallel GAs.

Finally the research can be extended by including more migration topologies, for instance more variants
(e.g. both directed and undirected) of traditional ones, as well as new ones: for example Watts-Strogatz networks or
other models of small-world and scale-free networks which can be found in papers like \cite{BarabasiAlbert02}. In
addition to that, more network parameters could be considered for identifying beneficial and unfavorable features for a
particular parallel algorithm.

\bibliographystyle{elsarticle-num}
\bibliography{bibliography}

\end{document}